\documentstyle[12pt]{article}
\date{}
\author{\it Valerii Dryuma\\[5mm]
{\it Institute of Mathematics and Informatics, AS RM,}\\[3mm]
{\it 5 Academiei Street, 2028 Kishinev, Moldavia}}
\title{  ON DUAL EQUATION IN THEORY \\[1mm]OF THE SECOND ORDER ODE's }

\newtheorem{pr}{Proposition}
\newtheorem{rem}{Remark}
\begin{document}
\maketitle

\begin{abstract}
\ \ \ \ We study the relations between  the
        second order nonlinear differential equations
        $$
        y''+a_{1}(x,y)y'^3+3a_{2}(x,y)y'^2+3a_{3}(x,y)y'+a_{4}(x,y)=0
        $$
        with arbitrary coefficients $a_{i}(x,y)$ and dual the second order
        nonlinear differential equations $$b''= g(a,b,b')$$
        with the function $g(a,b,b'=c)$ satisfying the  nonlinear partial
        differential equation
        $$
        g_{aacc}+2cg_{abcc}+2gg_{accc}+c^2g_{bbcc}+2cgg_{bccc}
        $$
        $$
        +g^2g_{cccc}+(g_a+cg_b)g_{ccc}-4g_{abc}-4cg_{bbc} -cg_{c}g_{bcc}
        $$
        $$
        -3gg_{bcc}-g_cg_{acc}+ 4g_cg_{bc}-3g_bg_{cc}+6g_{bb} =0\>.
        $$

\end{abstract}

\section{Introduction}

     The relation between the equations in form
\begin{equation}
  y''+a_{1}(x,y)y'^3+3a_{2}(x,y)y'^2+3a_{3}(x,y)y'+a_{4}(x,y)=0  \label{Lio}
\end{equation}
and
\begin{equation}
b''=g(a,b,b') \label{Dua1}
\end{equation}
 with function $g(a,b,b')$ satisfying  the p.d.e
$$
     g_{aacc}+2cg_{abcc}+2gg_{accc}+c^2g_{bbcc}+2cgg_{bccc}+
$$
\begin{equation}
    g^2g_{cccc}+(g_a+cg_b)g_{ccc}-4g_{abc}-4cg_{bbc} -cg_{c}g_{bcc}-
\end{equation}
$$
  3gg_{bcc}-g_cg_{acc}+ 4g_cg_{bc}-3g_bg_{cc}+6g_{bb} =0 \label{Dua2}.
$$
from geometrical point of view was studied by E.Cartan \cite{Cartan1}.

  In fact, according to the exp\-res\-sions on curvature of the
space of linear elements (x,y,y') connected with equation (1)
$$
\Omega^1_2=a[\omega^2 \wedge \omega^2_1]\,,\quad
\Omega^0_1=b[\omega^1 \wedge \omega^2]\,,\quad
\Omega^0_2=h[\omega^1 \wedge \omega^2]+k[\omega^2 \wedge \omega^2_1]\>.
$$
where:
$$
a=-\frac{1}{6}\frac{\partial^4 f}{\partial y'^4}\,,\quad h=\frac{\partial b}{\partial y'}\,, \quad
k=-\frac{\partial \mu}{\partial y'}-\frac{1}{6}\frac{\partial^2 f}{\partial^{2} y'}\frac{\partial^3 f}{\partial^{3} y'}\,,
$$
and
\begin{eqnarray*}
6b & = & f_{xxy'y'}+2y'f_{xyy'y'}+2ff_{xy'y'y'}+y'^2f_{yyy'y'}+2y'ff_{yy'y'y'} \\
   & + & f^2f_{y'y'y'y'}+(f_x+y'f_y)f_{y'y'y'}-4f_{xyy'}-4y'f_{yyy'} - y'f_{y'}f_{yy'y'}\\
   & - &  3ff_{yy'y'}-f_{y'}f_{xy'y'}+ 4f_{y'}f_{yy'}-3f_{y}f_{y'y'}+6f_{yy} \>.
\end{eqnarray*}
two types of equations by a natural
way are evolved: the first type from the con\-di\-tion $a =0$ and second type
from the condition $b =0$.

    The first condition $a=0$ lead to the the equation in form (1)
 and the second condition give us the equations (2) where the
function $g(a,b,b')$ satisfies the above p.d.e.~(3).

    From the elementary point of view the relation between both equations
(1) and (2) is a result of the special properties of their General
Integral $$ F(x,y,a,b)=0
 $$
which can be  considered as the equation of some 3-dim orbifold.

\section{Method of solution and reductions}

    The equation (2) forming dual pair with some equation (1)
    can be find from the solutions of the p.d.e.(3).

    For solutions of this type of equation we use the method of solution of the p.d.e.'s
     described first in \cite{Dr4}.

      To integrate the partial nonlinear first order differential equation
\begin{equation}\label{dryuma:eq1}
F(x,y,z,f_x,f_y,f_z,f_{xx},f_{xy},f_{xz},f_{yy},f_{yz},f_{xxx},f_{xyy},f_{xxy},..)=0
 \end{equation}
    can be applied a following approach.

      We use the following parametric presentation of the functions and variables
\begin{equation}\label{dryuma:eq2}
f(x,y,z)\rightarrow u(x,t,z),\quad y \rightarrow v(x,t,z),\quad
f_x\rightarrow u_x-\frac{u_t}{v_t}v_x,\]\[ f_z\rightarrow
u_z-\frac{u_t}{v_t}v_z,\quad f_y \rightarrow \frac{u_t}{v_t},
\quad f_{yy} \rightarrow \frac{(\frac{u_t}{v_t})_t}{v_t}, \quad
f_{xy} \rightarrow \frac{(u_x-\frac{u_t}{v_t}v_x)_t}{v_t},...
\end{equation}
where variable $t$ is considered as parameter.

  Remark that conditions of the type
   \[
   f_{xy}=f_{yx},\quad f_{xz}=f_{zx}...
   \]
are fulfilled at the such type of presentation.

  In result instead of equation (\ref{dryuma:eq1}) one get the
  relation between the new variables $u(x,t,z)$ and $v(x,t,z)$ and
  their partial derivatives
\begin{equation}\label{dryuma:eq3}
\Phi(u,v,u_x,u_z,u_t,v_x,v_z,v_t...)=0.
  \end{equation}

    In some cases the solution of such type of indefinite equation is more simple
    problem than solution of the equation (\ref{dryuma:eq1}).

    The equation (3) has many types of
reductions and the simplest of them are $$
 g=c^{\alpha}\omega[ac^{\alpha-1}],\quad g=c^{\alpha}\omega[bc^{\alpha-2}],
\quad g=c^{\alpha}\omega[ac^{\alpha-1},bc^{\alpha-2}], \quad
g=a^{-\alpha}\omega[ca^{\alpha-1}], $$ $$ \quad
g=b^{1-2\alpha}\omega[cb^{\alpha-1}], \quad g=a^{-1}\omega(c-b/a),
\quad g=a^{-3}\omega[b/a,b-ac],\quad
 g=a^{\beta/\alpha-2}\omega[b^{\alpha}/a^{\beta},
c^{\alpha}/a^{\beta-\alpha}]. $$

    For every type of reduction we can write corresponding equation
~(3) and then integrate it. Remark that the first examples of
solutions of equation (3) were obtained in [2-8].

\begin{pr}
 Equation ~(3) can be represent in form
\begin{eqnarray}
   g_{ac} +  gg_{cc} - g_{c}^{2}/2 + cg_{bc} -2 g_{b} = h(a,b,c),\\\nonumber
h_{ac}  +   gh_{cc} - g_{c}h_{c} + ch_{bc} -3h_{b} = 0.
\end{eqnarray}
\end{pr}

  So in standard name of variables we get
\begin{equation}\label{dryuma:eq5}
   f_{xy} +  ff_{yy} - f_{y}^{2}/2 + yf_{yz} -2 f_{z} = h(x,z,y),$$$$
h_{xy}  +   fh_{yy} - f_{y}h_{y} + yh_{xz} -3h_{z} = 0.
\end{equation}

\section{Two-dimensional short-cut $(x,y)$ equation}

        At the condition $h(x,z,y)=0$ we get the equation
\begin{equation}\label{dryuma:eq6}
   f_{xy} +  ff_{yy} - f_{y}^{2}/2 + yf_{yz} -2 f_{z} = 0.
\end{equation}

      In particular case $f(x,z,y)=f(x,y)$ the equation takes
      the form
\begin{equation}\label{dryuma:eq7}
   f_{xy} +  ff_{yy} - f_{y}^{2}/2 = 0.
\end{equation}

It was integrated in the \cite{Dr41} by the Legendre
transformation and another methods in more latest publications.

       We consider a new approach to integration of this equation.

     The change from the equation
(\ref{dryuma:eq7}) to the (\ref{dryuma:eq3}) lead to the relation
     between the functions $u(x,t)$ and $v(x,t)$ and their derivatives
\[
2\,\left ({\frac {\partial ^{2}}{\partial t\partial
x}}u(x,t)\right ) \left ({\frac {\partial }{\partial
t}}v(x,t)\right )^{2}-2\,\left ({ \frac {\partial ^{2}}{\partial
{t}^{2}}}u(x,t)\right )\left ({\frac {
\partial }{\partial x}}v(x,t)\right ){\frac {\partial }{\partial t}}v(
x,t)+\]\[+2\,\left ({\frac {\partial }{\partial t}}u(x,t)\right
)\left ({ \frac {\partial }{\partial x}}v(x,t)\right ){\frac
{\partial ^{2}}{
\partial {t}^{2}}}v(x,t)-2\,\left ({\frac {\partial }{\partial t}}u(x,
t)\right )\left ({\frac {\partial ^{2}}{\partial t\partial
x}}v(x,t) \right ){\frac {\partial }{\partial
t}}v(x,t)+\]\[+2\,u(x,t)\left ({\frac {
\partial ^{2}}{\partial {t}^{2}}}u(x,t)\right ){\frac {\partial }{
\partial t}}v(x,t)-2\,u(x,t)\left ({\frac {\partial }{\partial t}}u(x,
t)\right ){\frac {\partial ^{2}}{\partial {t}^{2}}}v(x,t)-\left ({
\frac {\partial }{\partial t}}u(x,t)\right )^{2}{\frac {\partial
}{
\partial t}}v(x,t)=0
\]

      The substitutions here of the form
    \[
    u(x,t)=t~ \partial_t\omega(x,t)-\omega(x,t),
\]
\[
v(x,t)=\partial_t\omega(x,t)
\]
 give us the linear p.d.e. equation
\begin{equation}\label{dryuma:eq8}
-2\,{\frac {\partial ^{2}}{\partial t\partial x}}\omega(x,t)+2\,t{
\frac {\partial }{\partial t}}\omega(x,t)-2\,\omega(x,t)-{t}^{2}{
\frac {\partial ^{2}}{\partial {t}^{2}}}\omega(x,t)=0.
\end{equation}

      The equation (\ref{dryuma:eq8}) is transformed into the form
\[
{\frac {\partial ^{2}}{\partial \eta\partial
\xi}}\omega(\xi,\eta)+4\, {\frac {{\frac {\partial }{\partial
\xi}}\omega(\xi,\eta)}{-\xi+\eta}} -2\,{\frac
{\omega(\xi,\eta)}{\left (-\xi+\eta\right )^{2}}}  =0
\]
with the change of variables
 \[\xi = x+2/t,\quad \eta = x
 \]
and can be integrated by the Laplace-method.

     In particular case the equation (\ref{dryuma:eq8}) admits the solution
     \[
    \omega(x,t)=-{\it \_C1}+4\,{\it \_C2}\,\ln (t)t+4\,{\it
\_C2}\,t+{\it \_C3}\,t+{\it \_C4}\,{t}^{2}+x\left ({\it
\_C1}\,t+{\it \_C2}\,{t}^{2} \right )
\]
with arbitrary parameters $\_Ci$ and
   elimination of the parameter $t$ from the relations
    \[
f(x,y)-t{\frac {\partial }{\partial t}}\omega(x,t)+\omega(x,t)
 =0
 \]
 and
 \[
 y-{\frac {\partial }{\partial t}}\omega(x,t)=0
\]
give us the solution   of the equation (\ref{dryuma:eq7}) taking
of of the form at the condition  $\_C4=0$
\[
f(x,y)=4\,{\it \_C2}\,\left ({\it LambertW}(1/2\,x{e^{-1/4\,{\frac
{-y+8\,{ \it \_C2}+{\it \_C3}+{\it \_C1}\,x}{{\it \_C2}}}}})\right
)^{2}{x}^{-1 }+\]\[+8\,{\it \_C2}\,{\it
LambertW}(1/2\,x{e^{-1/4\,{\frac {-y+8\,{\it \_C2}+{\it \_C3}+{\it
\_C1}\,x}{{\it \_C2}}}}}){x}^{-1}+{\it \_C1}.
\]

    So the equation
\begin{equation}\label{dryuma:eq9}
\frac{d^2 b}{da^2}=4\,{\it \_C2}\,\left ({\it
LambertW}(1/2\,a{e^{-1/4\,{\frac {-{\frac { d}{da}}b(a)+8\,{\it
\_C2}+{\it \_C3}+{\it \_C1}\,a}{{\it \_C2}}}}}) \right
)^{2}{a}^{-1}+\]\[+8\,{\it \_C2}\,{\it LambertW}(1/2\,a{e^{-1/4\,{
\frac {-{\frac {d}{da}}b(a)+8\,{\it \_C2}+{\it \_C3}+{\it
\_C1}\,a}{{ \it \_C2}}}}}){a}^{-1}+{\it \_C1}
 \end{equation}
is dual for the some equation in form (1).

   In fact General solution of the equation
   (\ref{dryuma:eq9}) is defined by the relation
\[b-8\,{\it \_C2}\,a+8\,x{\it \_C2}-{\it \_C3}\,a+{\it \_C3}\,x-1/2\,{
\it \_C1}\,{a}^{2}+1/2\,{\it \_C1}\,{x}^{2}-\]\[-4\,{\it
\_C2}\,a\ln (2)+4 \,{\it \_C2}\,x\ln (2)-4\,{\it \_C2}\,a\ln
(\left (-a+x\right )^{-1})- y(x)=0,
\]
and elimination of the variables $a,~b$ from here give us the
equation in form (1)
\begin{equation}\label{dryuma:eq10}
{\frac {d^{2}}{d{x}^{2}}}y(x)+\]\[+1/4\,{\frac {\left ({\frac
{d}{dx}}y(x) \right )^{2}}{x{\it \_C2}}}+1/4\,{\frac {\left
(-8\,{\it \_C2}\,\ln (2 )-2\,{\it \_C3}-2\,{\it \_C1}\,x-12\,{\it
\_C2}\right ){\frac {d}{dx}} y(x)}{x{\it \_C2}}}+\]\[+1/4\,{{\it
\_C1}}^{2}x+2\,{\it \_C1}\,{\it \_C2}\, \ln (2)+1/2\,{\it
\_C3}\,{\it \_C1}+2\,{\it \_C2}\,{\it \_C1}+\]\[+1/4\,{ \frac
{32\,{{\it \_C2}}^{2}+12\,{\it \_C2}\,{\it \_C3}+{{\it \_C3}}^{2
}+8\,{\it \_C3}\,{\it \_C2}\,\ln (2)+48\,{{\it \_C2}}^{2}\ln
(2)+16\,{ {\it \_C2}}^{2}\left (\ln (2)\right )^{2}}{x}}=0.
\end{equation}

    Remark that the equation (\ref{dryuma:eq10}) is the Rikkati
    equation with respect of variable $ z(x)=\frac{d y}{dx}$.

\section{Two-dimensional short-cut $(z,y)$ equation}

     The next example is the equation (\ref{dryuma:eq6}) at the
     conditions
     $h(x,z,y)=0$ and $f(x,z,y)=f(z,y)$
\begin{equation}\label{dryuma:eq11}
     ff_{yy} - f_{y}^{2}/2 + yf_{yz} -2 f_{z} = 0.
\end{equation}

For a such equation we get the relation
\[
2\,u(z,t)\left ({\frac {\partial ^{2}}{\partial
{t}^{2}}}u(z,t)\right ){\frac {\partial }{\partial
t}}v(z,t)-2\,u(z,t)\left ({\frac {
\partial }{\partial t}}u(z,t)\right ){\frac {\partial ^{2}}{\partial {
t}^{2}}}v(z,t)-\left ({\frac {\partial }{\partial t}}u(z,t)\right
)^{2 }{\frac {\partial }{\partial t}}v(z,t)+\]\[+2\,v(z,t)\left
({\frac {
\partial ^{2}}{\partial t\partial z}}u(z,t)\right )\left ({\frac {
\partial }{\partial t}}v(z,t)\right )^{2}-2\,v(z,t)\left ({\frac {
\partial ^{2}}{\partial {t}^{2}}}u(z,t)\right )\left ({\frac {
\partial }{\partial z}}v(z,t)\right ){\frac {\partial }{\partial t}}v(
z,t)+\]\[+2\,v(z,t)\left ({\frac {\partial }{\partial
t}}u(z,t)\right ) \left ({\frac {\partial }{\partial
z}}v(z,t)\right ){\frac {\partial ^ {2}}{\partial
{t}^{2}}}v(z,t)-2\,v(z,t)\left ({\frac {\partial }{
\partial t}}u(z,t)\right )\left ({\frac {\partial ^{2}}{\partial t
\partial z}}v(z,t)\right ){\frac {\partial }{\partial t}}v(z,t)-\]\[-4\,
\left ({\frac {\partial }{\partial z}}u(z,t)\right )\left ({\frac
{
\partial }{\partial t}}v(z,t)\right )^{3}+4\,\left ({\frac {\partial }
{\partial t}}u(z,t)\right )\left ({\frac {\partial }{\partial
z}}v(z,t )\right )\left ({\frac {\partial }{\partial
t}}v(z,t)\right )^{2}=0
\]
which is equivalent the p.d.e
\begin{equation}\label{dryuma:eq12}
-2\,\omega(z,t)+2\,t{\frac {\partial }{\partial
t}}\omega(z,t)-{t}^{2} {\frac {\partial ^{2}}{\partial
{t}^{2}}}\omega(z,t)-2\,\left ({\frac {\partial }{\partial
t}}\omega(z,t)\right ){\frac {\partial ^{2}}{
\partial t\partial z}}\omega(z,t)+\]\[+4\,\left ({\frac {\partial ^{2}}{
\partial {t}^{2}}}\omega(z,t)\right ){\frac {\partial }{\partial z}}
\omega(z,t)=0
 \end{equation}
  after the substitution
\[
    u(z,t)=t~ \partial_t\omega(z,t)-\omega(z,t),
\]
\[
v(z,t)=\partial_t\omega(z,t).
\]

   A simplest solution of the equation (\ref{dryuma:eq12}) can be find in form
   \[
\omega(z,t)=A(t)+z{t}^{2}
\]
 where the function $A(t)$ satisfies the linear equation
\[
-2\,A(t)-2\,t{\frac {d}{dt}}A(t)+3\,{t}^{2}{\frac
{d^{2}}{d{t}^{2}}}A( t)=0
\]
having the solution
\[
A(t)={\frac {{\it \_C1}}{\sqrt [3]{t}}}+{\it \_C2}\,{t}^{2}.
\]

     After inverse transformation in the case  $\_C2=0$ we find corresponding solution of
      the equation  (\ref{dryuma:eq11}) in implicit form
\[ -351918\,{y}^{3}{z}^{2}{{\it \_C1}}^{3}\left (f(z,y)\right
)^{2}+84672 \,{y}^{5}z{{\it \_C1}}^{3}f(z,y)-34992\,{y}^{2}\left
(f(z,y)\right )^{ 6}{z}^{2}+\]\[+8748\,{y}^{4}\left (f(z,y)\right
)^{5}z+46656\,\left (f(z,y) \right )^{7}{z}^{3}-729\,{y}^{6}\left
(f(z,y)\right )^{4}+518616\,y{z} ^{3}{{\it \_C1}}^{3}\left
(f(z,y)\right )^{3}+\]\[+823543\,{z}^{4}{{\it
\_C1}}^{6}-6912\,{y}^{7}{{\it \_C1}}^{3}=0.
\]

    So the second order ODE
    \[-351918\,\left ({\frac {d}{da}}b(a)\right )^{3}\left (b(a)\right )^{2}
{{\it \_C1}}^{3}\left ({\frac {d^{2}}{d{a}^{2}}}b(a)\right
)^{2}+84672 \,\left ({\frac {d}{da}}b(a)\right )^{5}b(a){{\it
\_C1}}^{3}{\frac {d^ {2}}{d{a}^{2}}}b(a)-\]\[-34992\,\left ({\frac
{d}{da}}b(a)\right )^{2} \left ({\frac
{d^{2}}{d{a}^{2}}}b(a)\right )^{6}\left (b(a)\right )^{2
}+8748\,\left ({\frac {d}{da}}b(a)\right )^{4}\left ({\frac
{d^{2}}{d{ a}^{2}}}b(a)\right )^{5}b(a)+\]\[+46656\,\left ({\frac
{d^{2}}{d{a}^{2}}}b( a)\right )^{7}\left (b(a)\right
)^{3}-729\,\left ({\frac {d}{da}}b(a) \right )^{6}\left ({\frac
{d^{2}}{d{a}^{2}}}b(a)\right )^{4}+\]\[+518616\, \left ({\frac
{d}{da}}b(a)\right )\left (b(a)\right )^{3}{{\it \_C1}}^ {3}\left
({\frac {d^{2}}{d{a}^{2}}}b(a)\right )^{3}+823543\,\left (b(a
)\right )^{4}{{\it \_C1}}^{6}-\]\[-6912\,\left ({\frac
{d}{da}}b(a)\right ) ^{7}{{\it \_C1}}^{3} =0
\]
is dual equation for the some equation of the form  (1).

   It can be reduced to the first order ODE
   \[
   -351918\,\left (h(b)\right )^{5}{b}^{2}{{\it \_C1}}^{3}\left ({\frac {
d}{db}}h(b)\right )^{2}+84672\,\left (h(b)\right )^{6}b{{\it
\_C1}}^{3 }{\frac {d}{db}}h(b)-\]\[-34992\,\left (h(b)\right
)^{8}\left ({\frac {d}{d b}}h(b)\right )^{6}{b}^{2}+8748\,\left
(h(b)\right )^{9}\left ({\frac {d}{db}}h(b)\right
)^{5}b+\]\[+46656\,\left ({\frac {d}{db}}h(b)\right )^{7 }\left
(h(b)\right )^{7}{b}^{3}-729\,\left (h(b)\right )^{10}\left ({
\frac {d}{db}}h(b)\right )^{4}+\]\[+518616\,\left (h(b)\right
)^{4}{b}^{3}{ {\it \_C1}}^{3}\left ({\frac {d}{db}}h(b)\right
)^{3}+823543\,{b}^{4}{ {\it \_C1}}^{6}-\]\[-6912\,\left
(h(b)\right )^{7}{{\it \_C1}}^{3}=0
\]
having singular solution
\[
h(b)={\frac {7}{108}}\,{108}^{5/7}\sqrt [7]{-{b}^{4}{{\it
\_C1}}^{3}}.
\]

     In general case $\_C1\neq0,\quad \_C2\neq0$
     we also get the first order ODE having  singular solution in the form
\[
\left ({\frac {d}{da}}b(a)\right )^{7}+{\frac
{823543}{11664}}\,{{\it \_C1}}^{3}{{\it \_C2}}^{4}+{\frac
{823543}{2916}}\,{{\it \_C1}}^{3}{{ \it \_C2}}^{3}b+\]\[+{\frac
{823543}{1944}}\,{{\it \_C1}}^{3}{{\it \_C2}}^{ 2}{b}^{2}+{\frac
{823543}{2916}}\,{{\it \_C1}}^{3}{\it \_C2}\,{b}^{3}+ {\frac
{823543}{11664}}\,{{\it \_C1}}^{3}{b}^{4}=0.
\]
which corresponds  the function determined from the equation
 \[
a+4\,\left (-1/12\,b(a)-1/12\,{\it \_C2}\right )\sqrt
[7]{-11664\,{ \frac {1}{{{\it \_C1}}^{3}\left ({\it
\_C2}+b(a)\right )^{4}}}}+{\it \_C3}=0 .
\]

    In more general case the solution of the equation
    (\ref{dryuma:eq12}) has the form
        \[
   \omega(z,t)=\left ({t}^{{\frac {k}{-1+4\,k}}}\right )^{4}{ \_C1}
\left ({t}^{\left (-1+4\,k\right )^{-1}}\right )^{-2}{t}^{-1}+{
\_C2}\,{t}^{2}+kz{t}^{2}
\]
where $k$ is essential parameter.

 With  help of the function $\omega(z,t)$ a
  large class of solutions of the equation (\ref{dryuma:eq11})
radically depending from the choice of parameter $k$ can be
produced.

\section{Two-dimensional full $(x,y)$- equation}

    In the case  $f(x,z,y)=f(x,y)$, ~$h(x,z,y)=h(x,y)\neq0$ from the system (\ref{dryuma:eq5}) we find
   full $(x,y)$ - equation
    \begin{equation}\label{dryuma:eq13}
{\frac {\partial ^{4}}{\partial y\partial {x}^{2}\partial
y}}f(x,y)+ \left ({\frac {\partial }{\partial x}}f(x,y)\right
){\frac {\partial ^ {3}}{\partial {y}^{3}}}f(x,y)+2\,f(x,y){\frac
{\partial ^{4}}{
\partial {y}^{2}\partial x\partial y}}f(x,y)+\left (f(x,y)\right )^{2}
{\frac {\partial ^{4}}{\partial {y}^{4}}}f(x,y)-\]\[-\left ({\frac
{
\partial }{\partial y}}f(x,y)\right ){\frac {\partial ^{3}}{\partial y
\partial x\partial y}}f(x,y)=0.
 \end{equation}

    This equation can be transformed into the form (\cite{Dr6})
    \begin{equation}\label{dryuma:eq14}
    \frac{\Omega_x}{\Omega_f}\Omega_{(xfff)}+ \frac{1}{\Omega_f} \Omega_{(xxff)}-1=0
    \end{equation}
with the help of  presentation
\[
y-\Omega(f(x,y),x)=0.
\]

     From the equation (\ref{dryuma:eq14}) we find that the
     function $\Omega_x(f,x)$ defined by the relation
\[
\Omega(f,x)=\frac{\partial \Lambda(x,f)}{\partial x}
\]
satisfies the equation
\[{\frac {\partial ^{2}}{\partial {f}^{2}}}\Lambda(x,f)=1/6\,\left ({\frac {
\partial }{\partial f}}\Lambda(x,f)\right )^{3}+\alpha(f)\left ({\frac {
\partial }{\partial f}}\Lambda(x,f)\right )^{2}+\beta(f){\frac {\partial }{
\partial f}}\Lambda(x,f)+\mu(f),
\]
with arbitrary coefficients.

   From solutions of this equation we can find the function $g(a,c)$ from the relation
\[c-\Omega(g(a,c),a)=0
\]
 and the equations forming dual pair with the equations
\begin{equation}\label{dryuma:eq15}
  y''+a_{1}(x)y'^3+3a_{2}(x)y'^2+3a_{3}(x)y'+a_{4}(x)=0
\end{equation}
 can be obtained by such a way.

   Remark that the equation (\ref{dryuma:eq15}) has the form of
   Abel's equation with respect the variable $z(x)=y'$.

    In the case of its solvability (Bernoulli and others) there are
   a lot possibilities to get an examples of of dual equation in explicit
   form.

    Let us consider the solutions  of the equation (\ref{dryuma:eq13}) in form
\[
Ay''^2+(By'^3+C y'^2+Ey'+F)y''+Hy'^6+K y'^5+Ly'^4+My'^3+N
y'^2+Py'+Q=0
\]
were the coefficients  depend   from the variable $x$
$A=A(x),~B=B(x)...$.

Joint consideration of the relation
\[
Af(x,y)^2+(By^3+C y^2+Ey+F)f(x,y)+Hy^6+K y^5+Ly^4+My^3+N
y^2+Py+Q=0
\]
and (\ref{dryuma:eq13}) lead to the conditions for determination
of the coefficients.

As example we find
\[
-26244\,{A}^{2}\left (f(x,y)\right )^{2}+\left
(18\,{x}^{3}{y}^{3}+ 2916\,Ax{y}^{2}\right
)f(x,y)+27\,{x}^{2}{y}^{4}+3888\,{y}^{3}A=0
\]
or
\[
-2916\,{A}^{2}\left ({\frac {d^{2}}{d{x}^{2}}}y(x)\right
)^{2}-324\,A \left ({\frac {d^{2}}{d{x}^{2}}}y(x)\right )\left
({\frac {d}{dx}}y(x) \right )y(x)+\]\[+3\,\left (y(x)\right
)^{2}\left ({\frac {d}{dx}}y(x) \right )^{2}-2\,\left (y(x)\right
)^{3}{\frac {d^{2}}{d{x}^{2}}}y(x)+ 432\,A\left ({\frac
{d}{dx}}y(x)\right )^{3}=0.
\]

From  General Integral \[ y{\it \_C1}\,{x}^{2}+2\,y{\it
\_C1}\,x{\it \_C2}+y{\it \_C1}\,{{\it \_C2}}^{2}+108\,A-108\,A{\it
\_C1}\,x-108\,A{\it \_C1}\,{\it \_C2}=0,
\]
we find dual equation
\[ 3\,{\frac {d}{da}}b(a)+\left ({\frac
{d}{da}}b(a)\right )^{3}{a}^{4}+2 \,\left ({\frac
{d^{2}}{d{a}^{2}}}b(a)\right )a=0.
\]

The equation
\[
b''=A/b^3
\]
has General Integral
\[
{b}^{2}-x\left (a-y\right )^{2}-{\frac {A}{x}}=0
\]

Corresponding dual equation looks as
\[
-A{x}^{6}\left ({\frac {d^{2}}{d{x}^{2}}}y(x)\right )^{2}+\left
(-6\, \left ({\frac {d}{dx}}y(x)\right ){x}^{5}A-2\,\left ({\frac
{d}{dx}}y( x)\right )^{3}{x}^{9}\right ){\frac
{d^{2}}{d{x}^{2}}}y(x)+{A}^{2}-\]\[-3\, {x}^{8}\left ({\frac
{d}{dx}}y(x)\right )^{4}-6\,{x}^{4}\left ({\frac
{d}{dx}}y(x)\right )^{2}A=0
\]

\section{Short-cut $(x,y,z)$ equation}

    The equation has the form
     \[
    {\frac {\partial ^{2}}{\partial x\partial y}}f(x,y,z)+f(x,y,z){\frac {
\partial ^{2}}{\partial {y}^{2}}}f(x,y,z)-1/2\,\left ({\frac {
\partial }{\partial y}}f(x,y,z)\right )^{2}+y{\frac {\partial ^{2}}{
\partial y\partial z}}f(x,y,z)-2\,{\frac {\partial }{\partial z}}f(x,y
,z)=0
\]

On rearrangement  we find the relation
\[
2\,\left ({\frac {\partial ^{2}}{\partial t\partial
x}}u(x,t,z)\right )\left ({\frac {\partial }{\partial
t}}v(x,t,z)\right )^{2}-2\,\left ( {\frac {\partial }{\partial
t}}u(x,t,z)\right )\left ({\frac {
\partial ^{2}}{\partial t\partial x}}v(x,t,z)\right ){\frac {\partial
}{\partial t}}v(x,t,z)-\]\[-2\,\left ({\frac {\partial }{\partial
x}}v(x,t, z)\right )\left ({\frac {\partial ^{2}}{\partial
{t}^{2}}}u(x,t,z) \right ){\frac {\partial }{\partial
t}}v(x,t,z)+2\,\left ({\frac {
\partial }{\partial x}}v(x,t,z)\right )\left ({\frac {\partial }{
\partial t}}u(x,t,z)\right ){\frac {\partial ^{2}}{\partial {t}^{2}}}v
(x,t,z)+\]\[+2\,u(x,t,z)\left ({\frac {\partial ^{2}}{\partial
{t}^{2}}}u(x ,t,z)\right ){\frac {\partial }{\partial
t}}v(x,t,z)-2\,u(x,t,z)\left ({\frac {\partial }{\partial
t}}u(x,t,z)\right ){\frac {\partial ^{2}} {\partial
{t}^{2}}}v(x,t,z)-\]\[-\left ({\frac {\partial }{\partial t}}u(x,
t,z)\right )^{2}{\frac {\partial }{\partial
t}}v(x,t,z)+2\,v(x,t,z) \left ({\frac {\partial ^{2}}{\partial
t\partial z}}u(x,t,z)\right ) \left ({\frac {\partial }{\partial
t}}v(x,t,z)\right )^{2}-\]\[-2\,v(x,t,z) \left ({\frac {\partial
}{\partial t}}v(x,t,z)\right )\left ({\frac {
\partial ^{2}}{\partial {t}^{2}}}u(x,t,z)\right ){\frac {\partial }{
\partial z}}v(x,t,z)-\]\[-2\,v(x,t,z)\left ({\frac {\partial }{\partial t}}
v(x,t,z)\right )\left ({\frac {\partial }{\partial
t}}u(x,t,z)\right ) {\frac {\partial ^{2}}{\partial t\partial
z}}v(x,t,z)+\]\[+2\,v(x,t,z) \left ({\frac {\partial
^{2}}{\partial {t}^{2}}}v(x,t,z)\right )\left ({\frac {\partial
}{\partial t}}u(x,t,z)\right ){\frac {\partial }{
\partial z}}v(x,t,z)-\]\[-4\,\left ({\frac {\partial }{\partial t}}v(x,t,z)
\right )^{3}{\frac {\partial }{\partial z}}u(x,t,z)+4\,\left
({\frac {
\partial }{\partial t}}v(x,t,z)\right )^{2}\left ({\frac {\partial }{
\partial t}}u(x,t,z)\right ){\frac {\partial }{\partial
z}}v(x,t,z)=0.
\]

 From here with the help of substitution
 \[
    u(x,t,z)=t~ \partial_t\omega(z,t)-\omega(z,t),
\]
\[
v(x,t,z)=\partial_t\omega(z,t)
\]
we get the equation
 \begin{equation}\label{dryuma:eq16}
 -2\,{\frac {\partial ^{2}}{\partial t\partial x}}\omega(x,t,z)+2\,t{
\frac {\partial }{\partial
t}}\omega(x,t,z)-2\,\omega(x,t,z)-{t}^{2}{ \frac {\partial
^{2}}{\partial {t}^{2}}}\omega(x,t,z)-\]\[-2\,\left ({ \frac
{\partial }{\partial t}}\omega(x,t,z)\right ){\frac {\partial ^{
2}}{\partial t\partial z}}\omega(x,t,z)+4\,\left ({\frac {\partial
^{2 }}{\partial {t}^{2}}}\omega(x,t,z)\right ){\frac {\partial
}{\partial z}}\omega(x,t,z)=0.
 \end{equation}

     We consider  the solutions of the equation (\ref{dryuma:eq16}) in form
     \[
     \omega(x,t,z)=A(x,t)+k z{t}^{2}
     \]
where $k$ is parameter.

 In result we get linear equation for the function
$A(x,t)$
\[\left (2\,t-4\,kt\right ){\frac {\partial }{\partial t}}A(x,t)+\left (
-{t}^{2}+4\,k{t}^{2}\right ){\frac {\partial ^{2}}{\partial
{t}^{2}}}A (x,t)-2\,{\frac {\partial ^{2}}{\partial t\partial
x}}A(x,t)-2\,A(x,t)=0.
\]

    It can be transformed into the form
\begin{equation}\label{dryuma:eq17}
4\,\left (3\,k-1\right )\left (-\xi+\eta\right ){\frac {\partial
}{
\partial \eta}}A(\xi,\eta)-\left ({\frac {\partial ^{2}}{\partial \eta
\partial \xi}}A(\xi,\eta)\right )\left (-\xi+\eta\right )^{2}\left (-1
+4\,k\right )-2\,A(\xi,\eta)  =0
 \end{equation}
 with help of the substitutions
\[\left \{\xi=x,\eta=-{\frac {x}{-1+4\,k}}+4\,{\frac {xk}{-1+4\,k}}-2\,{
\frac {1}{\left (-1+4\,k\right )t}}\right \} .
 \]

 In result the Laplace-equation
\begin{equation}\label{dryuma:eq18}
{\frac {\partial ^{2}}{\partial \eta\partial
\xi}}A(\xi,\eta)-{\frac { \left (12\,k-4\right ){\frac {\partial
}{\partial \eta}}A(\xi,\eta)}{ \left (-1+4\,k\right )\left
(-\xi+\eta\right )}}+2\,{\frac {A(\xi,\eta )}{\left
(-\xi+\eta\right )^{2}\left (-1+4\,k\right )}}  =0
\end{equation}
with the invariants
\[
H=-2\,{\frac {1}{\left (-\xi+\eta\right )^{2}\left (-1+4\,k\right
)}}
\]
and
\[
K=6\,{\frac {2\,k-1}{\left (-\xi+\eta\right )^{2}\left
(-1+4\,k\right )}}
\]
has been obtained.

    For a given case we have
    \[
    p={\frac {K}{H}}=-6\,k+3
    \]
    and
    \[
    q=\frac{\partial_ \xi \partial_ \eta (ln H)}{H}=-1+4\,k
\]
i.e. all invariants of the Laplace-sequence of the equation
(\ref{dryuma:eq16}) are in a fixed ratio (\cite{Oves}).

 In particular the condition
\[
H^N/H=1+\left (1-p\right )N-1/2\,qN\left (N+1\right )
=\]\[=1+\left (6\,k-2\right )N-1/2\,\left (-1+4\,k\right )N\left
(N+1\right)
\]
is fulfilled.

     From here we have
    \[
    N=2,\quad N=-\left (-1+4\,k\right )^{-1}.
\]

\section{Full $(x,y,z)$- equation}

   The construction of non trivial solutions of a full $(x,y,z)$- equation
\begin{equation}\label{dryuma:eq19}
{\frac {\partial ^{4}}{\partial y\partial {x}^{2}\partial
y}}f(x,y,z)+ \left ({\frac {\partial }{\partial x}}f(x,y,z)\right
){\frac {
\partial ^{3}}{\partial {y}^{3}}}f(x,y,z)+2\,f(x,y,z){\frac {\partial
^{4}}{\partial {y}^{2}\partial x\partial
y}}f(x,y,z)-\]\[-4\,{\frac {
\partial ^{3}}{\partial y\partial x\partial z}}f(x,y,z)+2\,y{\frac {
\partial ^{4}}{\partial {y}^{2}\partial x\partial z}}f(x,y,z)+\left (f
(x,y,z)\right )^{2}{\frac {\partial ^{4}}{\partial
{y}^{4}}}f(x,y,z)+\]\[+2 \,f(x,y,z)y{\frac {\partial
^{4}}{\partial {y}^{3}\partial z}}f(x,y,z) -\left ({\frac
{\partial }{\partial y}}f(x,y,z)\right ){\frac {
\partial ^{3}}{\partial y\partial x\partial y}}f(x,y,z)+\]\[+4\,\left ({
\frac {\partial }{\partial y}}f(x,y,z)\right ){\frac {\partial
^{2}}{
\partial y\partial z}}f(x,y,z)-\left ({\frac {\partial }{\partial y}}f
(x,y,z)\right )y{\frac {\partial ^{3}}{\partial {y}^{2}\partial
z}}f(x ,y,z)+\]\[+y\left ({\frac {\partial }{\partial
z}}f(x,y,z)\right ){\frac {
\partial ^{3}}{\partial {y}^{3}}}f(x,y,z)-4\,y{\frac {\partial ^{3}}{
\partial z\partial y\partial z}}f(x,y,z)+{y}^{2}{\frac {\partial ^{4}}
{\partial z\partial {y}^{2}\partial z}}f(x,y,z)-\]\[-3\,\left
({\frac {
\partial }{\partial z}}f(x,y,z)\right ){\frac {\partial ^{2}}{
\partial {y}^{2}}}f(x,y,z)-3\,f(x,y,z){\frac {\partial ^{3}}{\partial
{y}^{2}\partial z}}f(x,y,z)+\]\[+6\,{\frac {\partial
^{2}}{\partial {z}^{2} }}f(x,y,z)=0 \end{equation}
 may be interested to understanding of the properties of $3$-dim orbifolds
 \[
  F(x,y,a,b)=0
\]
 defined by the second order ODE's (\ref{Lio}).

Here we describe some approach to solution of this problem.

    With this aim we introduce the function $K(\tau,x,z)$ by
    definition
\[
y''=f(x,y,z)=\sqrt{1+y'^2}K(\tau,x,z)=\frac{K(\tau,x,z)}{\cos(\tau)^3},
\]
where
\[
\tau=\arctan(y')
\]

   Function $K(\tau,x,z)$ is the curvature along the curve.

   From the equation (\ref{dryuma:eq19}) we find the equation for the
   function $K(\tau,x,z)$
   \begin{equation}\label{dryuma:eq20}
   -4\,\left ({\frac {\partial ^{3}}{\partial z\partial \tau\partial z}}K
(\tau,x,z)\right )\sin(2\,\tau)+4\,\left ({\frac {\partial ^{3}}{
\partial x\partial \tau\partial x}}K(\tau,x,z)\right )\sin(2\,\tau)-6
\,\left ({\frac {\partial ^{2}}{\partial x\partial z}}K(\tau,x,z)
\right )\sin(2\,\tau)+\]\[+2\,\left ({\frac {\partial
^{4}}{\partial x
\partial {\tau}^{2}\partial z}}K(\tau,x,z)\right )\sin(2\,\tau)+9\,{
\frac {\partial ^{2}}{\partial {x}^{2}}}K(\tau,x,z)+{\frac
{\partial ^ {4}}{\partial x\partial {\tau}^{2}\partial
x}}K(\tau,x,z)+9\,{\frac {
\partial ^{2}}{\partial {z}^{2}}}K(\tau,x,z)+\]\[+{\frac {\partial ^{4}}{
\partial z\partial {\tau}^{2}\partial z}}K(\tau,x,z)-8\,\left ({\frac
{\partial ^{3}}{\partial x\partial \tau\partial
z}}K(\tau,x,z)\right ) \cos(2\,\tau)+20\,\left (K(\tau,x,z)\right
)^{2}{\frac {\partial ^{2}} {\partial
{\tau}^{2}}}K(\tau,x,z)+\]\[+2\,\left (K(\tau,x,z)\right )^{2}{
\frac {\partial ^{4}}{\partial {\tau}^{4}}}K(\tau,x,z)+18\,\left
(K( \tau,x,z)\right )^{3}-\left ({\frac {\partial ^{4}}{\partial z
\partial {\tau}^{2}\partial z}}K(\tau,x,z)\right )\cos(2\,\tau)+\]\[+3\,
\left ({\frac {\partial ^{2}}{\partial {z}^{2}}}K(\tau,x,z)\right
) \cos(2\,\tau)-3\,\left ({\frac {\partial ^{2}}{\partial
{x}^{2}}}K( \tau,x,z)\right )\cos(2\,\tau)+\left ({\frac {\partial
^{4}}{\partial x\partial {\tau}^{2}\partial x}}K(\tau,x,z)\right
)\cos(2\,\tau)-\]\[-8\, \left ({\frac {\partial }{\partial
\tau}}K(\tau,x,z)\right )\left ({ \frac {\partial ^{2}}{\partial
\tau\partial x}}K(\tau,x,z)\right )\sin (\tau)+8\,\left ({\frac
{\partial }{\partial \tau}}K(\tau,x,z)\right ) \cos(\tau){\frac
{\partial ^{2}}{\partial \tau\partial z}}K(\tau,x,z)-\]\[-
2\,\left ({\frac {\partial }{\partial \tau}}K(\tau,x,z)\right
)\left ( {\frac {\partial ^{3}}{\partial {\tau}^{2}\partial
z}}K(\tau,x,z) \right )\sin(\tau)+8\,\left ({\frac {\partial
}{\partial \tau}}K(\tau, x,z)\right )\left ({\frac {\partial
}{\partial z}}K(\tau,x,z)\right ) \sin(\tau)+\]\[+6\,\left ({\frac
{\partial }{\partial x}}K(\tau,x,z)\right )\left ({\frac {\partial
^{2}}{\partial {\tau}^{2}}}K(\tau,x,z)\right
)\sin(\tau)+4\,K(\tau,x,z)\sin(\tau){\frac {\partial
^{4}}{\partial { \tau}^{3}\partial z}}K(\tau,x,z)-\]\[-2\,\left
({\frac {\partial }{
\partial \tau}}K(\tau,x,z)\right )\cos(\tau){\frac {\partial ^{3}}{
\partial {\tau}^{2}\partial x}}K(\tau,x,z)+28\,K(\tau,x,z)\sin(\tau){
\frac {\partial ^{2}}{\partial \tau\partial
z}}K(\tau,x,z)+\]\[+36\,K(\tau, x,z)\sin(\tau){\frac {\partial
}{\partial x}}K(\tau,x,z)+6\,K(\tau,x,z )\left ({\frac {\partial
^{3}}{\partial {\tau}^{2}\partial x}}K(\tau,x ,z)\right
)\sin(\tau)+\]\[+28\,K(\tau,x,z)\left ({\frac {\partial ^{2}}{
\partial \tau\partial x}}K(\tau,x,z)\right )\cos(\tau)+2\,\left ({
\frac {\partial }{\partial z}}K(\tau,x,z)\right )\sin(\tau){\frac
{
\partial ^{3}}{\partial {\tau}^{3}}}K(\tau,x,z)+\]\[+2\,\left ({\frac {
\partial }{\partial x}}K(\tau,x,z)\right )\cos(\tau){\frac {\partial ^
{3}}{\partial {\tau}^{3}}}K(\tau,x,z)-6\,K(\tau,x,z)\left ({\frac
{
\partial ^{3}}{\partial {\tau}^{2}\partial z}}K(\tau,x,z)\right )\cos(
\tau)+\]\[+8\,\left ({\frac {\partial }{\partial
\tau}}K(\tau,x,z)\right ) \cos(\tau){\frac {\partial }{\partial
x}}K(\tau,x,z)-6\,\left ({\frac {\partial }{\partial
z}}K(\tau,x,z)\right )\left ({\frac {\partial ^{2 }}{\partial
{\tau}^{2}}}K(\tau,x,z)\right )\cos(\tau)-\]\[-36\,K(\tau,x,z)
\left ({\frac {\partial }{\partial z}}K(\tau,x,z)\right
)\cos(\tau)+4 \,K(\tau,x,z)\cos(\tau){\frac {\partial
^{4}}{\partial {\tau}^{3}
\partial x}}K(\tau,x,z)=0.
\end{equation}

     A simplest solutions of the equation (\ref{dryuma:eq20}) has the form
     \begin{equation}\label{dryuma:eq21}
     K(\tau,x,z)=-\left ({\frac {\partial }{\partial x}}U(x,z)\right )\sin(
\tau)+\left ({\frac {\partial }{\partial z}}U(x,z)\right
)\cos(\tau) \end{equation}
 where
\[-6\,{\frac {\partial ^{3}}{\partial {x}^{3}}}U(x,z)-6\,{\frac {
\partial ^{3}}{\partial z\partial x\partial z}}U(x,z)+12\,\left ({
\frac {\partial ^{2}}{\partial {z}^{2}}}U(x,z)\right ){\frac {
\partial }{\partial x}}U(x,z)+12\,\left ({\frac {\partial ^{2}}{
\partial {x}^{2}}}U(x,z)\right ){\frac {\partial }{\partial
x}}U(x,z)=0
\]
and
\[-12\,\left ({\frac {\partial ^{2}}{\partial {x}^{2}}}U(x,z)\right ){
\frac {\partial }{\partial z}}U(x,z)-12\,\left ({\frac {\partial
}{
\partial z}}U(x,z)\right ){\frac {\partial ^{2}}{\partial {z}^{2}}}U(x
,z)+6\,{\frac {\partial ^{3}}{\partial {z}^{3}}}U(x,z)+6\,{\frac {
\partial ^{3}}{\partial {x}^{2}\partial z}}U(x,z)=0.
\]

  It follows that the function $U(x,z)$ is solution of the equation
  \[
  U_{xx}+U_{zz}=\exp(2U)
  \]
and condition (\ref{dryuma:eq21}) corresponds the equation
\[
z''=(U_z-z'U_x)(1+z'^2)
\]

     This equation is cubic on the first derivative and this case
     corresponds the projectively flat pair of the second order ODE's.

     As an illustration of nontrivial example can be considered the
     function
     \[
     K(\tau,x,z)=-\]\[=-{\frac {\left (\cos(\tau)\right )^{3}\left (2\,\tan(\tau)
{x}^{3}\left (1+\left (\tan(\tau)\right )^{2}{x}^{6}\right
)+2\,\left (1+\left (\tan(\tau)\right )^{2}{x}^{6}\right
)^{3/2}+3\,\tan(\tau){x} ^{3}\right )}{{x}^{4}}}
\]
which correspond the second order ODE
     \begin{equation}\label{dryuma:eq22}
z''=-{\frac {2\,z'{x}^{3}\left (1+{z'}^{2}{x}^{6}\right )+2\,\left
(1+{z'}^ {2}{x}^{6}\right )^{3/2}+3\,z'{x}^{3}}{{x}^{4}}}.
\end{equation}

    Equation (\ref{dryuma:eq22}) is dual equation for
 the some of the second order ODE cubic on the first derivative.

     It enters into the composition of the equations
     \[
     b''=g(a,b,b')=\frac{A(b'a^{\mu-1})}{a^{\mu}},
     \]
where the function $A(b'a^{\mu-1})=A(\xi)$ satisfies the equation
\[
(A+(\mu-1)\xi)^2A^{IV}+3(\mu-2)(A+(\mu-1)\xi)A^{III}+(2-\mu)A^{I}A^{II}+(\mu^2-5\mu+6)A^{II}=0.
\]

 This solution was considered here as an example of solution of the full
$f(x,y,z)$ - equation.

     In a most general case the solution of the equation
     (\ref{dryuma:eq20}) can be considered in the form
     \[
     K(\tau,x,z)=\frac{\sum_{n=0}^{\infty} (A_n(x,z)\sin(\tau n)+B_n(x,z)\cos(\tau
     n)+C_n(x,z))}{\cos(\tau)^3}.
     \]

\begin{rem}
The system (\ref{dryuma:eq5}) after the change of variables and
function according to the rule
\[
g(x,b',b)=b'\phi(x, \ln(b),b/b')
\]
takes the form
\[2\,{\frac {\partial }{\partial x}}\phi(x,\eta,\xi)+2\,{\frac {
\partial ^{2}}{\partial \eta\partial x}}\phi(x,\eta,\xi)-2\,\left ({
\frac {\partial ^{2}}{\partial x\partial
\xi}}\phi(x,\eta,\xi)\right ) \xi+2\,\phi(x,\eta,\xi){\frac
{\partial ^{2}}{\partial {\eta}^{2}}}
\phi(x,\eta,\xi)-\]\[-4\,\phi(x,\eta,\xi)\left ({\frac {\partial
^{2}}{
\partial \eta\partial \xi}}\phi(x,\eta,\xi)\right )\xi+2\,\phi(x,\eta,
\xi)\left ({\frac {\partial ^{2}}{\partial
{\xi}^{2}}}\phi(x,\eta,\xi) \right ){\xi}^{2}-\left
(\phi(x,\eta,\xi)\right )^{2}+\]\[+2\,\phi(x,\eta, \xi)\left
({\frac {\partial }{\partial \xi}}\phi(x,\eta,\xi)\right )
\xi-\left ({\frac {\partial }{\partial
\eta}}\phi(x,\eta,\xi)\right )^ {2}+2\,\left ({\frac {\partial
}{\partial \eta}}\phi(x,\eta,\xi) \right )\left ({\frac {\partial
}{\partial \xi}}\phi(x,\eta,\xi) \right )\xi-\]\[-\left ({\frac
{\partial }{\partial \xi}}\phi(x,\eta,\xi) \right
)^{2}{\xi}^{2}+2\,{\frac {\partial ^{2}}{\partial \eta\partial
\xi}}\phi(x,\eta,\xi)-2\,\left ({\frac {\partial ^{2}}{\partial
{\xi}^ {2}}}\phi(x,\eta,\xi)\right )\xi-4\,{\frac {\partial
}{\partial \xi}} \phi(x,\eta,\xi) =2\kappa(x,\eta,\xi)
\]
\[
{\frac {\partial ^{2}}{\partial \eta\partial
x}}\kappa(x,\eta,\xi)- \left ({\frac {\partial ^{2}}{\partial
x\partial \xi}}\kappa(x,\eta, \xi)\right
)\xi+\phi(x,\eta,\xi){\frac {\partial ^{2}}{\partial {\eta}
^{2}}}\kappa(x,\eta,\xi)-2\,\phi(x,\eta,\xi)\left ({\frac
{\partial ^{ 2}}{\partial \eta\partial
\xi}}\kappa(x,\eta,\xi)\right )\xi-\]\[-2\,\phi(x ,\eta,\xi){\frac
{\partial }{\partial \eta}}\kappa(x,\eta,\xi)+\phi(x,
\eta,\xi)\left ({\frac {\partial ^{2}}{\partial
{\xi}^{2}}}\kappa(x, \eta,\xi)\right
){\xi}^{2}+3\,\phi(x,\eta,\xi)\left ({\frac {\partial }{\partial
\xi}}\kappa(x,\eta,\xi)\right )\xi-\]\[-\left ({\frac {\partial
}{\partial \eta}}\phi(x,\eta,\xi)\right ){\frac {\partial
}{\partial \eta}}\kappa(x,\eta,\xi)+\left ({\frac {\partial
}{\partial \eta}}\phi (x,\eta,\xi)\right )\left ({\frac {\partial
}{\partial \xi}}\kappa(x, \eta,\xi)\right )\xi+\left ({\frac
{\partial }{\partial \xi}}\phi(x, \eta,\xi)\right )\xi\,{\frac
{\partial }{\partial \eta}}\kappa(x,\eta, \xi)-\]\[-\left ({\frac
{\partial }{\partial \xi}}\phi(x,\eta,\xi)\right ){ \xi}^{2}{\frac
{\partial }{\partial \eta}}\kappa(x,\eta,\xi)+{\frac {
\partial ^{2}}{\partial \eta\partial \xi}}\kappa(x,\eta,\xi)-\left ({
\frac {\partial ^{2}}{\partial
{\eta}^{2}}}\kappa(x,\eta,\xi)\right ) \xi-4\,{\frac {\partial
}{\partial \eta}}\kappa(x,\eta,\xi)=0,
\]
where
\[
\eta=\ln(b'),\quad \xi=b/b'.
\]
\end{rem}

    This form of equation can be of used for construction of a new examples
    of dual equations.

    In more general case exists the reduction in the form
    \[
    f=y^k\omega(xy^{(k-1)},\ln(y),xy^{(k-2)})
    \]

\section{Examples}

      Here we discuss  possibility to reception  of a new examples
      of dual equations.

      Let
      \[
      \phi(x,y,y')=C
      \]
be the first integral of the second order ODE
\[
y''=f(x,y,y').
\]

   Then from the relation
   \[
   \phi_x+y'\phi_{y}+f(x,y,y')\phi_{y'}=0
   \]
we find
\begin{equation}\label{dryuma:eq23}
f(x,y,y')=-\frac{\phi_x+y'\phi_{y}}{\phi_{y'}}.
 \end{equation}

     After substitution of this expression into the relation
     (\ref{dryuma:eq5}) we get the equation to determination of the
      function $\phi(x,y,y')$.

      Let us consider an examples.

     Substitution of the expression (\ref{dryuma:eq23}) into the
     (\ref{dryuma:eq6}) lead to the equation
\begin{equation}\label{dryuma:eq24}
2\,\left ({\frac {d}{d\eta}}A(\eta)\right )^{2}\left ({\frac
{d^{2}}{d {\eta}^{2}}}A(\eta)\right )A(\eta)-2\,\left ({\frac
{d}{d\eta}}A(\eta) \right )\left ({\frac
{d^{3}}{d{\eta}^{3}}}A(\eta)\right )\left (A( \eta)\right
)^{2}-3\,\left ({\frac {d}{d\eta}}A(\eta)\right )^{4}+\]\[+3\,
\left ({\frac {d^{2}}{d{\eta}^{2}}}A(\eta)\right )\left (A(\eta)
\right )^{2}=0 \end{equation}
 on the function
 \[
     \phi(x,y,z)=A\left({\frac {y}{z}}\right){x}^{-1}
,\quad \eta=\frac{y}{z}.\]

    The solution of the equation (\ref{dryuma:eq24})
is
 \[
A(\eta)={\frac {{\it LambertW}(1/2\,{e^{1/4\,{\it \_C2}\,{\it
\_C1}}}{ e^{1/4\,\eta\,{\it \_C1}}}{e^{-1}}){\it \_C3}}{{\it
LambertW}(1/2\,{e^ {1/4\,{\it \_C2}\,{\it
\_C1}}}{e^{1/4\,\eta\,{\it \_C1}}}{e^{-1}})+1}}
 \]
or
\[A({\frac {y}{z}})={\it LambertW}(1/2\,{e^{1/4\,{\frac {y{\it \_C1}}{z}
}}}{e^{-1}}){\it \_C3}\left ({\it LambertW}(1/2\,{e^{1/4\,{\frac
{y{ \it \_C1}}{z}}}}{e^{-1}})+1\right )^{-1}.
\]

   Using this expression it is possible to find the second order ODE
   \[
 \left ({\frac {d^{2}}{d{x}^{2}}}y(x)\right )y(x)x{\it \_C1}-4\,\left (
y(x)\right )^{2}\left ({\it LambertW}(1/2\,{e^{-1/4\,{\frac
{-\left ({ \frac {d}{dx}}y(x)\right ){\it
\_C1}+4\,y(x)}{y(x)}}}})\right )^{2}-\]\[-8 \,\left (y(x)\right
)^{2}{\it LambertW}(1/2\,{e^{-1/4\,{\frac {-\left ({\frac
{d}{dx}}y(x)\right ){\it \_C1}+4\,y(x)}{y(x)}}}})-4\,\left (y(
x)\right )^{2}-\left ({\frac {d}{dx}}y(x)\right )^{2}{\it \_C1}\,x
=0
\]
with General Integral
\begin{equation}
\label{dryuma:eq25}
 y-\left (-{\frac {a}{x-a}}\right
)^{-4\,a}{16}^{x}\left (-{\frac {x}{x -a}}\right )^{4\,x}b\left
(-2\,x+2\,a\right )^{-4\,a}=0.
\end{equation}

     From here is followed that the equation
     \[
     b''=g(a,b,b')
     \]
     must be cubic on the first derivative $b'$.

         Substitution of the expression (\ref{dryuma:eq23}) into the
     full equation (\ref{dryuma:eq19}) lead to the equation on the
     function $\phi(x,y,z)$ having the solution in form
     \[
     \phi(x,y,z)=A\left({\frac {y}{z}}\right){x}^{-1}
,\quad \eta=\frac{y}{z},
\]
where the function $A(\eta)$ satisfies the condition
\[ -\left
(A(\eta)\right )^{3}\left ({\frac {d}{d\eta}}A(\eta)\right )^{3
}{\frac {d^{5}}{d{\eta}^{5}}}A(\eta)+4\,\left (A(\eta)\right )^{2}
\left ({\frac {d}{d\eta}}A(\eta)\right )^{2}\left ({\frac
{d^{2}}{d{ \eta}^{2}}}A(\eta)\right )^{3}-2\,A(\eta)\left ({\frac
{d}{d\eta}}A( \eta)\right )^{4}\left ({\frac
{d^{2}}{d{\eta}^{2}}}A(\eta)\right )^{2 }-\]\[-36\,\left
(A(\eta)\right )^{3}\left ({\frac {d}{d\eta}}A(\eta) \right )\left
({\frac {d^{2}}{d{\eta}^{2}}}A(\eta)\right )^{2}{\frac {
d^{3}}{d{\eta}^{3}}}A(\eta)+8\,\left (A(\eta)\right )^{3}\left ({
\frac {d}{d\eta}}A(\eta)\right )^{2}\left ({\frac
{d^{2}}{d{\eta}^{2}} }A(\eta)\right ){\frac
{d^{4}}{d{\eta}^{4}}}A(\eta)-\]\[-3\,\left ({\frac {
d}{d\eta}}A(\eta)\right )^{6}{\frac
{d^{2}}{d{\eta}^{2}}}A(\eta)+24\, \left (A(\eta)\right )^{3}\left
({\frac {d^{2}}{d{\eta}^{2}}}A(\eta) \right )^{4}+6\,\left
(A(\eta)\right )^{3}\left ({\frac {d}{d\eta}}A( \eta)\right
)^{2}\left ({\frac {d^{3}}{d{\eta}^{3}}}A(\eta)\right )^{2
}+\]\[+3\,A(\eta)\left ({\frac {d}{d\eta}}A(\eta)\right
)^{5}{\frac {d^{3}} {d{\eta}^{3}}}A(\eta)-3\,\left (A(\eta)\right
)^{2}\left ({\frac {d}{d \eta}}A(\eta)\right )^{3}\left ({\frac
{d^{2}}{d{\eta}^{2}}}A(\eta) \right ){\frac
{d^{3}}{d{\eta}^{3}}}A(\eta)=0
\]
which is generalization of the equation (\ref{dryuma:eq24}).

    Its solution can be presented in form
    \[
    \eta=\int \!{\frac {h(g)}{g{e^{\int \!h(g){dg}+{\it \_C2}}}}}{dg}+{
\it \_C1} ,
\]
\[
A(\eta)={e^{\int \!{\frac {h(g)}{g}}{dg}+{\it \_C3}}} ,
\]
where the function $h(g)$ satisfies the equation
\[
-2\,{\frac {\left ({g}^{2}-2\,g+1\right )\left (h(g)\right
)^{3}}{{g}^ {2}}}-{\frac {\left (4\,g-3\right )\left (h(g)\right
)^{2}}{{g}^{2}}}+ {\frac {\left (3\,{g}^{2}{\frac
{d}{dg}}h(g)-3\,\left ({\frac {d}{dg}} h(g)\right )g-1\right
)h(g)}{{g}^{2}}}+\]\[+{\frac {\left ({\frac {d^{2}}{d
{g}^{2}}}h(g)\right )g+4\,{\frac {d}{dg}}h(g)}{g}}-3\,{\frac
{\left ({ \frac {d}{dg}}h(g)\right )^{2}}{h(g)}} =0.
\]

Its solution in turn is expressed trough the solution of the Abel
equation
\[
{\frac {d}{d{\it a}}}{\it b}({\it a})=-{\it a}\,\left (2\,{{ \it
a}}^{2}+7\,{\it a}+6\right )\left ({\it b}({\it a})\right
)^{3}+\left (-3\,{\it a}-7\right )\left ({\it b}({\it a})\right
)^{2}-3\,{\frac {{\it b}({\it a})}{{\it a}}}
\]
having elementary particular solutions.

  In explicit form we get
  \[
  {h}(g)=-{\it a}\,{e^{\int \!{\it b}({\it a}){d{
\it a}}+{\it \_C1}}}+{\it a}\,\left ({e^{\int \!{\it b}({\it a
}){d{\it a}}+{\it \_C1}}}\right )^{2}
\]
and
\[
{g}={\frac {-1+{e^{\int \!{\it b}({\it a}){d{\it a}}+{\it
\_C1}}}}{{e^{\int \!{\it b}({\it a}){d{\it a}}+{\it \_C1}}}}}.
\]

\section{Acknowledgement}

     This research was partially supported by the Grant 06.01 CRF of HCSTD
     ASM and author is thanking the RFBR for stimulation of this
     activity.

\end{document}